\documentclass[twocolumn,showpacs,preprintnumbers,amsmath,amssymb, showkeys]{revtex4}

\usepackage{graphicx}
\usepackage{dcolumn}
\usepackage{bm}
\usepackage{epsf}

\begin{document}


\title{The influence of tunneling matrix elements modification due to on-site Coulomb interaction on local tunneling conductivity}

\author{V.\,N.\,Mantsevich}
 \altaffiliation{vmantsev@spmlab.phys.msu.ru}
\author{N.\,S.\,Maslova}%
 \email{spm@spmlab.phys.msu.ru}
\affiliation{%
 Moscow State University, Department of  Physics,
119991 Moscow, Russia
}%

\date{\today }
5 pages, 2 figures
\begin{abstract}
Interplay between changes of energy levels and tunneling amplitudes
caused by localized electrons on-site Coulomb interaction depending
on non-equilibrium electron filling numbers is analyzed. Specific
features of local tunneling conductivity spectra for different
positions of localized states energy relative to the Fermi level
have been investigated by means of self-consistent mean field
approximation in the presence of non-equilibrium effects. The
conditions when modifications of tunneling transfer amplitude due to
changes of electron filling numbers in the presence of on-site
Coulomb interaction should be taken into account in tunneling
conductivity spectra have been revealed.
\end{abstract}

\maketitle

\section{Introduction}

Recently great attention is paid to both experimental and
theoretical investigations of kinetic processes in nanostructures,
in connection with possibility of ultra small size electronic
devices fabrication. Comparison between experimental results and
theoretical models provide information about tunneling mechanisms in
nanoscale junctions due to the presence of localized states
\cite{Arseyev}. Localized states formed by impurities at surface and
interfaces of semiconductors lead to strong modification of local
electronic structure and influence the behavior of tunneling
characteristics due to on-site Coulomb interaction of the sample and
tip conduction electrons with charged localized states
\cite{Dombrowski,Sullivan,Mahieu}. Most of the experiments are
carried out with the help of scanning tunneling
microscopy/spectroscopy technique
\cite{Dombrowski,Inglesfield,Panov} and theoretical calculations
usually deals with Green's functions formalism \cite{Hofstetter,
Mantsevich,Arseyev,Mantsevich_1} or ab-initio calculaions
\cite{Qian}.

It was predicted theoretically and proved experimentally that
on-site Coulomb repulsion of localized electrons leads to the
changing of energy values considerably even for deep impurity levels
\cite{Arseyev,Mantsevich}. It was found that taking into account
Coulomb interaction of conduction electrons with non-equilibrium
localized charges result in non-trivial behavior of tunneling
characteristics in the case of STM metallic tip positioned above the
impurity atom \cite{Hofstetter,Madhavan,Arseyev_1} and also leads to
formation of peculiarities in local tunneling conductivity analyzed
apart from the impurity \cite{Madhavan, Mantsevich}. But in all the
previous calculation tunneling transfer rates considered as a
constant values and influence of the impurity's non-equilibrium
electron filling numbers on tunneling transfer amplitudes due to
Coulomb correlations was neglected
\cite{Arseyev_1,Panov,Mantsevich,Hofstetter}. However, this effect
is a problem of great interest because influence rate depends on the
impurity type and solving of this problem gives an opportunity to
initialize impurity type with the help of STM technique. Another
question of great interest is what effect dominates in tunneling
contact: influence of Coulomb interaction on-site or on hopping
matrix elements. Correct investigation of this problem shows the way
for not only qualitative but also for quantitative theoretical
analysis of experimentally measured local tunneling conductivity,
but it is necessary to choose the correct starting approximation for
each type of impurity atom depending on the localized state energy
level position relative to the Fermi level. Previous works
demonstrate that for localized states with energy deep below the
Fermi level Hubbard-I model \cite{Hubbard} suites well and for
localized states with energy in the vicinity of Fermi level mean
field mixed valence approximation \cite{Anderson} can be applied.

So in this work we present detailed investigation of local tunneling
conductivity in vicinity of an impurity paying special attention to
the tunneling transfer amplitude dependence from non-equilibrium
electron filling numbers due to Coulomb correlations. On-site
Coulomb interaction will be also taken into account in our model.
Development of theoretical description suggested in
\cite{Mantsevich} and based only on on-site Coulomb interaction
should clarify the role of tunneling transfer amplitudes
modification due to the influence of impurity's non-equilibrium
electron filling numbers in formation of peculiar features observed
experimentally in local tunneling conductivity. We performed
calculations for two extreme cases when tunneling takes place
through the impurity state with deep energy level and shallow energy
level. We applied modified mean field self-consistent approach for
the impurity state with shallow energy level and in the case of
impurity with deep energy level modified Hubbard-{I} model was used.
We also compared our results with the calculations based on
non-modified models based only on on-site Coulomb interaction. Both
approaches were analyzed with the use of the Keldysh fomalism
\cite{Keldysh}.

\section{The suggested model and main results}
We shall analyze tunneling between the semiconductor surface and
metallic STM tip in the presence of Coulomb interaction for
impurities with deep and shallow energy levels in comparison with
Fermi level. Special attention will be payed to the impurity atom
non-equilibrium electron filling numbers influence on the values of
tunneling transfer amplitudes.
    The standard model system semiconductor-impurity atom-metallic
tip with constant tunneling transfer rates can be described by the
Hamiltonian: $\Hat{H}$:

$$\Hat{H}=\Hat{H}_{0}+\Hat{H}_{imp}+\Hat{H}_{tun}$$
\begin{eqnarray}
&\Hat{H}_{0}&=\sum_{k\sigma}\varepsilon_{k}c_{k\sigma}^{+}c_{k\sigma}+\sum_{p\sigma}\varepsilon_{p}c_{p\sigma}^{+}c_{p\sigma}+h.c.\nonumber\\
&\Hat{H}_{tun}&=\sum_{kd\sigma}t_{kd} c_{k\sigma}^{+}c_{d\sigma}+\sum_{pd\sigma}t_{pd} c_{d\sigma}^{+}c_{p\sigma}+h.c.\nonumber\\
&\Hat{H}_{imp}&=\sum_{d\sigma}\varepsilon_{d}c_{d\sigma}^{+}c_{d\sigma}+Un_{d\sigma}n_{d-\sigma}\
\label{1}
\end{eqnarray}

Indices $k$ and $p$ label the states in the left (semiconductor) and
right (tip) lead, respectively. The index $d$ indicates that
impurity electron operator is involved. $U$ is the on-site Coulomb
repulsion of localized electrons,
$n_{d\sigma}=c_{d\sigma}^{+}c_{d\sigma}$, $c_{d\sigma}$ destroys
impurity electron with spin $\sigma$. $\Hat{H}_{0}$ describes
conduction electrons of continuous spectrum states in the leads of
tunneling contact. $\Hat{H}_{tun}$ describes tunneling transitions
between the tunneling contact leads through the impurity state.
$\Hat{H}_{imp}$ corresponds to the electrons in the localized state
formed by the impurity atom.

We shall start from the case of impurity state with energy level
situated deep below the Fermi level. In the presence of Coulomb
interaction tunneling through deep impurity energy level can be
described with the help of Hubbard-{I} model
\cite{Hubbard}(Hamiltonian of this model has the form \ref {1}).
Taking into account Coulomb interaction in this model leads to
formation of two well separated impurity energy levels
$\varepsilon_d$ and $\varepsilon_d+U$ instead of one initial level
$\varepsilon_d$. It is reasonable to use approximation in which the
strongest interaction of the considered model - the on-site Coulomb
repulsion $U$ - is included in $G^{0R\sigma}_{dd}(\omega)$. So we
can write down expression for $G^{0R\sigma}_{dd}(\omega)$ supposing
tunneling transfer amplitudes to be independent from localized state
non-equilibrium filling numbers:

\begin{eqnarray}
G^{0R\sigma}_{dd}(\omega)=\frac{1}{\omega-\varepsilon_{d}-\Sigma(\omega)}\nonumber\\
\Sigma(\omega)=\frac{n_{d-\sigma}U(\omega-\varepsilon_{d})}{\omega-\varepsilon_{d}-(1-n_{d-\sigma})U}
\label{2}
\end{eqnarray}

Using system of Dyson equations one can get expression for impurity
retarded Green's function.

\begin{eqnarray}
G^{R\sigma}_{dd}(\omega)=\frac{1}{\omega-\varepsilon_{d}-\Sigma(\omega)-i(\gamma_{kd}+\gamma_{pd})}
\label{3}
\end{eqnarray}

Where $\sum_{p}t_{pd}^{2}ImG_{pp}^{0R}=\gamma_{pd}$ and
$\sum_{k}t_{kd}^{2}ImG_{kk}^{0R}=\gamma_{kd}$. Expression for local
tunneling conductivity in the case of constant transfer amplitudes
was found in \cite{Maslova} and has the form:

\begin{eqnarray}
\frac{dI}{dV}(\omega)&=\frac{\gamma_{kd}\gamma_{pd}}{\gamma_{kd}+\gamma_{pd}}Im
G_{dd}^{R\sigma}(\omega) \label{4}
\end{eqnarray}

If one wants now to analyze modification of tunneling transfer
amplitudes due to the dependence from impurity's non-equilibrium
electron filling numbers, it is necessary to rewrite part
$\Hat{H}_{tun}$ of the initial Hamiltonian \ref{1} which corresponds
to tunneling transitions between the tunneling contact leads through
the impurity state. Due to Coulomb repulsion energy level position
depends on electron filling numbers. Impurity energy increases with
the increasing of filling numbers and it leads to the decreasing of
tunneling amplitudes between tunneling contact leads through the
impurity energy level. So higher energy level corresponds to the
smaller value of tunneling transfer amplitude and tunneling
amplitude now depends on electron presence or absence on the energy
level. Initial Hamiltonian \ref{1} can be rewritten in the following
way:

\begin{eqnarray}
&\Hat{H}_{tun}&=\sum_{k,d\sigma}t_{kd}^{eff}c_{k\sigma}^{+}c_{d\sigma}+\sum_{p,d\sigma}t_{pd}^{eff}c_{d\sigma}^{+}c_{p\sigma}+h.c.\
\label{5}
\end{eqnarray}

where expressions for $t_{kd}^{eff}$ and $t_{pd}^{eff}$ have the
form:

\begin{eqnarray}
t_{kd}^{eff}=(T_{1}(1-<n_{d-\sigma}>)+T_{2}<n_{d-\sigma}>)\nonumber\\
t_{pd}^{eff}=(t_{1}(1-<n_{d-\sigma}>)+t_{2}<n_{d-\sigma}>)\label{5.1}
\end{eqnarray}

When impurity energy level is free from electrons tunneling
transitions from semiconductor continuous spectrum states to STM tip
through impurity atom are described by transfer amplitudes $T_{1}$
and $t_{1}$. If impurity energy level is occupied by electron with
spin tunneling transitions between continuous spectrum states
through impurity state can be described by tunneling transfer
amplitudes $T_{2}$ and $t_{2}$. Due to this assumption initial
Hubbard-$I$ model \cite{Hubbard} can be modified and expression for
retarded Green's function will have the form:

\begin{eqnarray}
G^{R\sigma}_{dd}(\omega)=\frac{1}{\omega-\varepsilon_{d}-\Sigma(\omega)+i(\gamma_{kd}^{eff}+\gamma_{pd}^{eff})}\label{6}
\end{eqnarray}

In this case relaxation rates can be written in the following way
$\gamma_{kd}^{eff}=\sum_{k}t_{kd}^{eff}ImG_{kk}^{0R}$ and
$\gamma_{pd}^{eff}=\sum_{p}t_{pd}^{eff}ImG_{pp}^{0R}$. Expression
for local tunneling conductivity than will have the form:

\begin{eqnarray}
\frac{dI}{dV}(\omega)&=\frac{\gamma_{kd}^{eff}\gamma_{pd}^{eff}}{\gamma_{kd}^{eff}+\gamma_{pd}^{eff}}Im
G_{dd}^{R\sigma}\label{7}
\end{eqnarray}

Calculating expression for $ImG^{R\sigma}_{dd}(\omega)$ requires
solution of self-consistent system of equations. Two of them are
equations for $ImG^{R\pm\sigma}_{dd}(\omega)$, obtained from
expression \ref{6}, and three equations determine impurity atom
non-equilibrium electron filling numbers:

\begin{eqnarray}
n_{d\mp\sigma}=\frac{-1}{\pi}\int d\omega
n_{d\mp\sigma}(\omega)ImG^{R\pm\sigma}_{dd}(\omega,n_{d\pm\sigma})\nonumber\\
n_{d\sigma}(\omega)=n_{d-\sigma}(\omega)=\frac{\gamma_{kd}^{eff}n^{0}_k(\omega)+\gamma_{pd}^{eff}n^{0}_p(\omega)}{\gamma_{kd}^{eff}+\gamma_{pd}^{eff}}\nonumber\\
\label{8}
\end{eqnarray}

where $n^{0}_k(\omega)$ and $n^{0}_p(\omega)$ are equilibrium
filling numbers in the tunneling contact leads.

\begin{figure} [h]
\includegraphics[width=70mm]{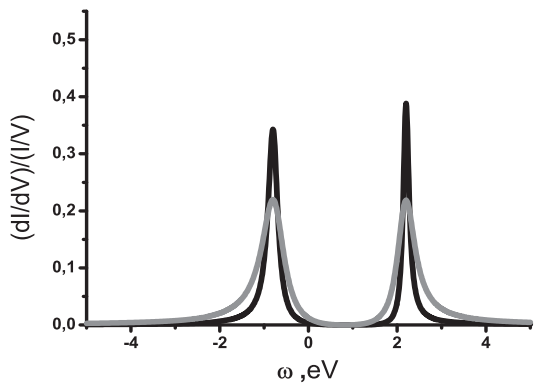}%
\caption{Local tunneling conductivity as a function of applied bias
voltage calculated for the impurity atom with deep energy level.
Gray line corresponds to the case when tunneling transfer amplitudes
depend on impurity filling numbers $t_1=0,20$, $t_2=0,29$,
$T_1=0,30$, $T_2=0,35$. Black line corresponds to the case when
tunneling transfer amplitudes have constant values and only on-site
Coulomb interaction is taken into account $t_{kd}=0,35$,
$t_{pd}=0,29$. For all the figures parameters values
$\varepsilon_{d}=-0,80$, $U=3$ are the same.}
\end{figure}

Fig.1 shows tunneling conductivity as a function of applied bias
voltage for impurity with energy level deep below Fermi level. It
means that relation between energy level value and energy level
width due to the tunneling processes is large than unit. Tunneling
conductivity calculated with the use of non-modified model
Hubbard-{I} (Hamiltonian of this model has the form \ref{1}) is
shown by the black line and results obtained with the use of
modified model (expression \ref{5} for $\Hat{H}_{tun}$) where
dependence of tunneling transfer amplitudes from impurity
non-equilibrium electron filling numbers is taken into account are
depicted by the gray line. Tunneling conductivity calculated above
the impurity demonstrates resonant peaks when applied bias voltage
is equal to the initial position of impurity energy level
$(\omega=\varepsilon_{d})$ and when applied bias voltage is equal to
the initial position of impurity energy level shifted on the value
of Coulomb potential  $(\omega=\varepsilon_{d}+U)$ both for taking
(Fig.~1 gray line) and not taking into account dependence of
tunneling transfer amplitudes from impurity non-equilibrium electron
filling numbers (Fig.~1 black line).

Modification of tunneling transfer amplitudes by non-equilibrium
electron filling numbers results strongly in significant decreasing
of resonant peaks amplitudes in comparison with the calculations
performed with the use of Hubbard-{I} model for constant values of
tunneling transfer amplitudes (Fig.~1 black line). Resonant
peculiarities also demonstrate considerable width changing. When
tunneling transfer amplitudes modification is taken into account
resonant peaks width increases in comparison with calculations
performed with the use of non-modified Hubbard-$I$ model.

So one can clearly see that Coulomb interaction influence on hopping
matrix elements in addition to on-site Coulomb interaction strongly
modify local tunneling conductivity measured above impurity with
energy deep below Fermi level. Taking into account both effects
leads to significant amplitude and width corrections of
peculiarities in modified conductivity.

Now let us analyze tunneling through the impurity state with energy
level in the vicinity of Fermi level. It means that relation between
energy level value and energy level width due to the tunneling
processes is about unit. In this case Coulomb interaction effects
accompanied by the assumption that non-equilibrium electron filling
numbers modify tunneling transfer amplitudes can be described with
the use of modified mean-field mixed valence approximation
\cite{Meir,Sivan} and consequently Hamiltonian depends on average
electron filling numbers. So one has to rewrite again part
$\Hat{H}_{tun}$ of the initial Hamiltonian \ref{1} in the form
\ref{5}. More over impurity energy level position depends on Coulomb
interaction of the non-equilibrium electron density. A new impurity
energy level position is a result of initial energy level position
shift on the value $U\cdot<n_{d-\sigma}>$. Non-equilibrium impurity
atom filling numbers $n_{d-\sigma}$ must satisfy self-consistency
condition \ref{8} where $\gamma_{kd}^{eff}$ and $\gamma_{pd}^{eff}$
are determined by expressions \ref{5.1} and
$G^{R\sigma}_{dd}(\omega)$ in the mean field approximation has the
following form:

\begin{eqnarray}
G^{R\sigma}_{dd}(\omega)=\frac{1}{\omega-\varepsilon_{d}-U<n_{d-\sigma}>+i(\gamma_{kd}^{eff}+\gamma_{pd}^{eff})}\
\label{9}
\end{eqnarray}

After determining new position of impurity energy level we can
calculate expressions for $ImG^{R\sigma}_{dd}(\omega)$ from \ref{9}
and finally evaluate local tunneling conductivity for shallow
impurity taking into account Coulomb interaction effects and
tunneling transfer amplitudes modification influenced by the
impurity non-equilibrium electron filling numbers with the use of
expression \ref{7}.

\begin{figure}[h]
\centering
\includegraphics[width=70mm]{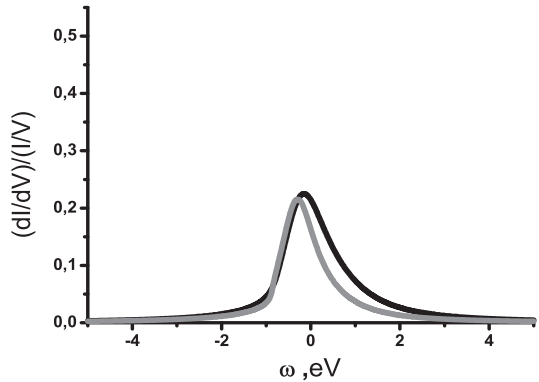}%
\caption{Local tunneling conductivity as a function of applied bias
voltage calculated for the impurity atom with shallow energy level.
Gray line corresponds to the case when tunneling transfer amplitudes
depend on impurity filling numbers $t_1=0,40$, $t_2=0,49$,
$T_1=0,60$, $T_2=0,68$. Black line corresponds to the case when
tunneling transfer amplitudes have constant values and only on-site
Coulomb interaction is taken into account $t_{kd}=0,68$,
$t_{pd}=0,49$. For all the figures parameters values
$\varepsilon_{d}=-0,80$, $U=3$ are the same.}
\end{figure}

Calculation results are presented on Fig.2. Tunneling conductivity
as a function of applied bias voltage demonstrates that taking into
account modification of tunneling transfer amplitudes due to the
influence of impurity's non-equilibrium electron filling numbers
(Fig.2 gray line) in addition to Coulomb interaction leads to
important changes in comparison with the case of constant tunneling
rates and on-site Coulomb interaction(Fig.2 black line). Decreasing
of resonant peak width and a negligible decreasing of the tunneling
conductivity amplitude takes place for modified self-consistent mean
field approximation due to the influence of impurity's
non-equilibrium electron filling numbers on tunneling transfer
amplitudes and consequently on relaxation rates. Describing
tunneling through the impurity state with energy level in the
vicinity of Fermi level it is necessary to mention that width of
peculiarities in tunneling conductivity is determined by the value
of relaxation rates $\gamma_{kd}^{eff}$ and $\gamma_{pd}^{eff}$.
Continuous energy shift of the resonant peak to the higher values of
applied bias voltage exists mostly due to on-site Coulomb
interaction of localized electrons
($U<n_{d-\sigma}>$$>$$\gamma^{eff}$) and can be seen both for taking
and not taking into account tunneling matrix elements modification.
It is clearly evident that resonant peak shift for modified
tunneling amplitudes is smaller than in the case of constant
tunneling amplitudes also because of relaxation rates modification.

\section{Conclusion}
In this work we have analyzed the role of tunneling transfer
amplitudes modification caused by the shift of energy level with
changing of non-equilibrium electron filling numbers in the presence
of on-cite Coulomb interaction in formation of anomalous features in
local tunneling conductivity. The influence of this effect on local
tunneling conductivity have been analyzed for different location of
localized states energy relative to Fermi level. We have studied two
extreme cases when tunneling takes place through the impurity state
with energy level position deep below Fermi level and through
impurity state with energy level in the vicinity of Fermi level by
means of modified Hubbard-{I} model and modified self-consistent
mixed valence mean field approximation correspondingly.

For impurity state with energy level deep below the Fermi level
modification of tunneling transfer amplitudes in addition to on-site
Coulomb interaction results in significant changing of resonant
peaks amplitudes and width in comparison with non-modified
Hubbard-{I} model. Resonant peculiarities demonstrate strong
amplitudes decreasing. We also observed considerable increasing of
resonant peculiarities width.

Impurity state with energy in the vicinity of Fermi level also
demonstrates significant changes in local tunneling conductivity in
comparison with the case of constant tunneling rates due to
modification of tunneling transfer amplitudes in addition to on-site
Coulomb interaction. We observed width changing of resonant peak,
slight decreasing of the tunneling conductivity amplitude and
smaller resonant peak shift to higher value of applied bias voltage.

Obtained results give us possibility to conclude that on-site
Coulomb interaction strongly influence local tunneling conductivity
measured in the vicinity of impurity atom and Coulomb interaction
influence on hopping matrix elements leads to corrections of
peculiarities shape, amplitude and width in modified conductivity.

This work was supported by RFBR grants and by the National Grants
for technical regulation and metrology.


\pagebreak


\begin{thebibliography}{99}
\bibitem{Arseyev}
P.I.~Arseyev, N.S.~Maslova et al., {\em JETP Letters}, {\bf 68},
299, (1998)

\bibitem{Dombrowski}
R.~Dombrowski, C.~Wittneven, M.~Morgenstern et al., {\em Appl. Phys.
A}, {\bf 66}, S203-S206, (1998)

\bibitem{Sullivan}
J.~Sullivan, G.~Boishin, L.~Whitman et al., {\em Phys. Rev. B}, {\bf
68}, 235324, (2003)

\bibitem{Mahieu}
G.~Mahieu, B.~Grandidier, D.~Deresmes et al., {\em Phys. Rev.
Lett.}, {\bf 94}, 026407, (2005)

\bibitem{Inglesfield}
J.~Inglesfield, M.~Boon, S.~Crampin, {\em Condens. Matter}, {\bf
12}, L489-L496, (2000)

\bibitem{Panov}
N.S.~Maslova, V.I.~Panov, S.I.~Oreshkin et al., {\em JETP Letters },
{\bf 67}, 130, (1998)

\bibitem{Hofstetter}
W.~Hofstetter, J.~Konig, H.~Schoeller, {\em Phys. Rev. Letters},
{\bf 87}, 156803, (2001)

\bibitem{Mantsevich}
V.N.~Mantsevich, N.S.~Maslova, {\em Solid State Communications},
{\bf 150}, 2072, (2010)

\bibitem{Mantsevich_1}
V.N.~Mantsevich, N.S.~Maslova, {\em JETP Letters}, {\bf 91}, 150,
(2010)

\bibitem{Qian}
M.~Qian, M.~Gothelid, B.~Johnsson, {\em Phys. Rev. B}, {\bf 66},
155326, (2002)

\bibitem{Madhavan}
V.~Madhavan, W.~Chen, T. Jamneala et al., {\em Science}, {\bf 280},
567, (1998)

\bibitem{Arseyev_1}
P.I.~Arseyev, N.S.~Maslova et al., {\em ZheTF}, {\bf 121}, 225,
(2002)
\bibitem{Hubbard}
J.~Hubbard, {\em Proc. Roy. Soc.}, {\bf 276}, 238, (1963)

\bibitem{Anderson}
P.W.~Anderson, {\em Phys. Rev.}, {\bf 115}, 1439, (1959)

\bibitem {Keldysh}
L.V.~Keldysh, {\em Sov. Phys JETP}, {\bf 20}, 1018, (1964)

\bibitem{Mantsevich_2}
V.N.~Mantsevich, N.S.~Maslova, {\em JETP Letters}, {\bf 89}, 609,
(2009)

\bibitem{Maslova}
P.I.~Arseyev, N.S.~Maslova, {\em ZETF}, {\bf 102}, 575, (1992)

\bibitem{Meir}
Y.~Meir, N.~Wingreen et al., {\em Phys. Rev. Letters}, {\bf 70},
2601, (1993)

\bibitem{Sivan}
N.~Sivan, N.~Wingreen, {\em Phys. Rev. B}, {\bf 54}, 11622, (1996)
\end{thebibliography}
\end{document}